\documentstyle[12pt,a4]{article}
\begin{document}

\title{\vspace{-1cm}
 \hfill {\small THEF-NYM-93.14} \vspace{-.3cm}\\
\hfill{\small gr-qc/9403053} \vspace{-.3cm}\\
 \hfill {\small March 1994} \\[.5cm]
\Large \bf Post-Newtonian Limits and Gravitational Radiation
from the Superstring Effective Action}
\author{ Alexandros A. Kehagias
        \thanks{Supported by CEC. Contract No. ERBCHBGCT920197}
\\      Institute of Theoretical Physics
\\    University of Nijmegen, Toernooiveld 1,
\\     6525 Nijmegen, The Netherlands}
\date{}
\maketitle
\begin{abstract}
\begin{sloppypar}
\normalsize

 We consider the gravitational sector of the  superstring effective
action with axion-matter couplings.  The field equations are developed
in the post-Newtonian scheme and approximate solutions for a spinning
point mass and a  cosmic string are presented.  Furthermore, assuming
vanishing axion mass and  vanishing potential for the dilaton, we
consider the gravitational radiation in the leading $O(\alpha'^0)$
order.  We find that the total luminosity of a radiative source has
monopole and dipole components besides the standard quadrupole one.
These components may possibly be checked in binary systems.

\end{sloppypar} \end{abstract}

\newpage

\section{Introduction}

 After the first attemps to formulate a consistent quantum theory of
 gravity \cite{Q}, it became apparent, and strongly supplied by the
non-renormalizability of general relativity \cite{NR}, that the
effective action of such a theory would involve, besides the Einstein
term, also terms in higher powers in the curvature \cite{C}. These
terms will determine the short range (high-frequency) behavior of the
theory
 while, at large distances, the theory will be classical gravity as
 described by the  (low-frequency)  Einstein term. However, the
incorporation of such high-power curvature terms in the action does not
lead to the resolution of the renormalization problem. The reason is
that the equations one obtains are of order higher than two. As a
result, the corresponding theory  has, inevitably,  indefinite energy
at the
 classical level and indefinite metric Hilbert space with ghosts at the
quantum level.

Today, string theory \cite{ST} seems to be the best candidate for a
consistent formulation of quantum gravity and, in addition, it has the
prospects to unify  all interactions. However, the greatest challenge
for string theory, as well as of any theory, is to make contact with
experiment, and even more, to make predictions. On the other hand, all
particle properties
 are  originated from the particular solution of the string difficult,
in view also of the absence of any classification, to extract results
 and to make predictions. In this case, one may look for common
features of the solutions and  such a  feature is the  presence of
tensor exitations (metric).  Thus, gravitation is a universal aspect of
any closed critical string theory and string-generated gravity is,
potentialy, a promising region  for  tests and confrontation of string
theory with experimental and observational data.

The graviton excitations in any closed critical string theory is a
member of a multiplet which contains, besides graviton, also the
massless modes of the scalar dilaton and the Kalb-Ramond  antisymmetric
two-form field \cite{ST}.
 The leading term in the effective action, in a conformal frame,
 consists of  the Einstein term and the kinetic terms for the other
 fields. In the next-to-leading order in the $\alpha'$-expansion and at
 the string tree level, the universal part of the effective  action in
any four-dimensional heterotic superstring model takes the form
\cite{S1,LZ,S2} \begin{eqnarray}
 I & =& \int d^4x\sqrt{-g}  (\frac{1}{2\kappa^2}R-\frac{1}{2}
\partial_\mu \Phi \partial^\nu \Phi - 6e^{-2\sqrt{2}\kappa\Phi}
H_{\mu\nu\rho}H^{\mu\nu\rho} \nonumber \\ & &  +
\frac{\alpha'}{16\kappa^2} e^{-\sqrt{2}\kappa
\Phi}(R^2_{GB}-F_{\mu\nu}F^{\mu\nu}) + \lambda e^{-\sqrt{2}\kappa \Phi}
H_{\kappa\lambda\rho} J^5_\mu \varepsilon^{\kappa\lambda\rho\mu} +
{\cal L}_{matter} ).  \end{eqnarray} where
$\varepsilon^{\mu\nu\rho\sigma}$ is the totally antisymmetric symbol
(with $\varepsilon_{0123}=+1$) and \begin{equation} R_{GB}^2=R^{\mu \nu
\rho \sigma}R_{\mu \nu \rho \sigma}-4R^{\mu \nu}R_{\mu \nu} +R^2,
\end{equation} is the Gauss-Bonnet tensor. This term has previously
been considered in higher-dimensional pure gravity theories \cite{L}
and has also been employed in the Kaluza-Klein framework \cite{KK}. The
new feature in the action (1) is that we have incorporated the coupling
of the antisymmetric three form $H_{\mu\nu\rho}$ field to the axial
vector $J^5_\mu=\overline{\psi} \gamma^5 \gamma_\mu \psi$ with, in our
notation, $\lambda=\kappa/12$ \cite{LZ}.  (More correctly, it is the
expectation value of the axial vector that $H_{\mu\nu\rho}$ should be
coupled.) Moreover, we have collectivelly denote by ${\cal L}_{matter}$
all field  contributions  which will be served as the source for
gravity, since the variation of this term will provide the ``effective"
energy-momentum tensor.

The antisymmetric $H_{\mu\nu\rho}$-field is given by \begin{equation}
H_{\mu\nu\rho}=\partial_{[\mu} B_{\nu\rho ]}+ \frac{\alpha'}{16\kappa}
(\omega^{3L}_{\mu\nu\rho}-\omega^{3\Upsilon}_{\mu\nu\rho})
\end{equation} where $B_{\mu\nu}$ is the antisymmetric component of the
graviton multiplet and \[
\partial_{[\mu}B_{\nu\rho]}=\frac{1}{3!}(\partial_\mu B_{\nu\rho}+
\partial_\rho B_{\mu\nu}+\partial_\nu B_{\rho\mu}).  \] Moreover, the
Lorentz and gauge Cern-Simons three-forms $\omega^{3L}, \omega^{3Y}$
are given, with $Tr$ denoting trace over suppressed Lorentz and gauge
indices, by \begin{eqnarray} \omega^{3L} & = & Tr(\omega \wedge
R-\frac{2}{3} \omega \wedge \omega \wedge \omega) \nonumber \\
\omega^{3\Upsilon} & = & Tr(A \wedge F-\frac{2}{3} A\wedge A \wedge A)
\end{eqnarray} with $A$ the gauge field and $\omega$ the
spin-connection. Usually, in the absence of sources, one may proceed
further and write, at least locally, the three form with components
$H_{\mu\nu\rho}$ as the dual of an exact one form, namely as
\begin{equation} H_{\mu\nu\rho}=\frac{1}{\sqrt{-g}}
\varepsilon_{\mu\nu\rho\kappa} \partial^\kappa a \end{equation} where
$a$ is the axion field. Solution for the axion \cite{Ax1,Ax2,Ax3} and
the dilaton \cite{D} have largely been studied, the former in
particular in a Kerr-Newmann background. However, it is not possible
when sources are present, to express the H-field as in  eq(5) although
it may  always be written as the dual of a non-exact one-form.

The major reason for studing the string-generated gravity is that, as
we have mentioned, the gravitational sector  is a generic feature of
any string theory and thus, its study
 will improve our understanding in a model independent way. Moreover,
keeping in mind that general relativity is an experimentally tested
theory \cite{Will}, perhaps there is also room for similar test in
string theory \cite{K1}. This is why an extensive study of the theory
is needed which will lead, hopefully, to predictions of effects not
present in general relativity.  For such a purpose, one does not need
to go through all the details of the theory. All predictions of general
relativity for example, have been extracted in some approximation
scheme, namely, in the post-Newtonian or the weak-field approximation.

In the following two chapters, we are dealing with the post-Newtonian
approximation method which is appropriate for slow moving bound
systems. We  also examine in chapter 4 the gravitational radiation in
the weak-field approximation.

\section{The post-Newtonian approximation}

 The  action (1) leads to nonlinear field equations which cannot be
 solved in general.  Exact solutions can  be found in special cases and
a guide to such a search, as in general relativity, will be the
isometries of spacetime. However, in view of the complexity of the
field equations, it is important some systematic approximation methods
to be developed, so that the main properties of the system under study
to become  apparent. There are
 two such methods, the post-Newtonian and the weak-field approximation
\cite{W1,F}.  Here we will deal with the former case which is based on
the assumption of slow moving gravitational bound systems
 and let us recall some well established results of this method based
mainly in \cite{W1}.

 We assume that there exist an expansion of the metric in inverse
 powers of the speed of light for slow moving gravitational bound
systems. Since this speed
 has been fixed here to one, the post-Newtonian approximation
 corresponds to an expansion of the metric tensor in powers of a small
 quantity $\epsilon$ which may be taken to be a typical velocity of the
 system. Thus we expect an expansion of the form\footnote {The
 conventions we adapt here is that of ref\cite{H}, namely, +2 signature
 for the metric and Riemann tensor defined by
${R^\mu}_{\nu\lambda\kappa}=+\partial_
\lambda{\Gamma^\mu}_{\nu\kappa}-\partial_\kappa{\Gamma^\mu}_{\lambda\nu}+
{\Gamma^\mu}_{\lambda\rho}{\Gamma^\rho}_{\kappa\nu}-{\Gamma^\mu}_{\kappa\rho}
{\Gamma^\rho}_{\lambda\nu}$} \[ g_{00}=-1+ {^{(2)}g}_{00}+
{^{(4)}g}_{00}+...  \] \[ g_{ij}=\delta_{ij}+{^{(2)}g}_{ij}+
{^{(4)}g}_{ij}+...  \] \begin{equation} g_{0i}= {^{(3)}g}_{0i}+
{^{(5)}g}_{0i}+..., \end{equation} where $^{(n)}g_{\mu\nu}$ is of order
$\epsilon^n$.  Greek indices run from 0 to 3, while Latin ones from 1
to 3.  The inverse metric has also an expansion which may be written as
\[ g^{00}=-1+ {^{(2)}g}^{00}+ {^{(4)}g}^{00}+...  \] \[
g^{ij}=\delta^{ij}+ {^{(2)}g}^{ij}+ {^{(4)}g}^{ij}+...  \]
\begin{equation} g^{0i}= {^{(3)}g}^{0i}+ {^{(5)}g}^{0i}+...,
\end{equation} with \[ {^{(2)}g}^{00}=- {^{(2)}g}_{00},
{^{(2)}g}^{ij}=- {^{(2)}g}_{ij},..., \] as one may easily verify using
\[ g_{\mu\nu}g^{\mu\lambda}=\delta_{\nu}^\lambda.  \]

This  expansion of the metric gives rise to a similar expansion for the
connection. In fact, since the connection is expressed in terms of the
metric as \[ {\Gamma^\lambda}_{\mu\nu}=\frac{1}{2}g^{\lambda\kappa}(
\partial_\nu g_{\kappa\mu}+\partial_\mu g_{\kappa\nu}- \partial_\kappa
g_{\mu\nu}) , \] we find that $\Gamma$-symbols with even number of
zeros (e.g. $ {\Gamma^i}_{00}$) may be written as \[
{\Gamma^\mu}_{\nu\lambda}= {^{(2)}\Gamma^\mu}_{\nu\lambda}+
{^{(4)}\Gamma^\mu}_{\nu\lambda}+... \ , \] while those  with an odd
number  (e.g.  ${\Gamma^i}_{j0}$)  as \[ {\Gamma^\mu}_{\nu\lambda}=
{^{(3)}\Gamma^\mu}_{\nu\lambda}+ {^{(5)}\Gamma^\mu}_{\nu\lambda}+...
\ .  \] This follows from the assumption that the expansions in powers
of $\epsilon$ holds in the near zone $r<t$ and thus spatial derivatives
are of order $O(\epsilon^0)$ while time ones are of $O(\epsilon^1)$
order.

The first  terms in the above  expansion  for the connection may
explicitly be written as \begin{eqnarray} {^{(2)}\Gamma^i}_{00} & = &
-\frac{1}{2}\partial_i{^{(2)}g}_{00} \nonumber \\ {^{(2)}\Gamma^0}_{i0}
& = & {^{(2)}\Gamma^i}_{00} \nonumber \\ {^{(2)}\Gamma^i}_{jk} & = &
\frac{1}{2}( \partial_k{^{(2)}g}_{ij} + \partial_j{^{(2)}g}_{ik} -
\partial_i{^{(2)}g}_{jk}) \nonumber \\ {^{(3)}\Gamma^0}_{00} & = &
-\frac{1}{2}\partial_t{^{(2)}g}_{00} \nonumber \\ {^{(3)}\Gamma^i}_{j0}
& = & \frac{1}{2}( \partial_t{^{(2)}g}_{ij} + \partial_j{^{(2)}g}_{i0}
- \partial_i{^{(2)}g}_{j0}) \nonumber \\ {^{(4)}\Gamma^i}_{00} & = &
-\frac{1}{2} {^{(2)}g}^{ij}(\partial_j{^{(2)}g}_{00} +
\partial_t{^{(3)}g}_{i0} -\frac{1}{2}\partial_i{^{(4)}g}_{00})
\nonumber \\ {^{(4)}\Gamma^i}_{jk} & = & \frac{1}{2}(
\partial_k{^{(4)}g}_{ij} + \partial_j{^{(4)}g}_{ik} -
\partial_i{^{(4)}g}_{jk}) \nonumber \\ {^{(4)}\Gamma^0}_{i0} & = &
{^{(4)}\Gamma^i}_{00} \end{eqnarray}

It is  straightforward now to verify that the Riemann tensor, as
defined in footnote (1), has an expansion in even powers in $\epsilon$
for components with even number of zeros, (e.g. ${R^0}_{m0n}$) and in
odd powers in $\epsilon$ for odd ones (e.g. ${R^0}_{mnk}$). Explicitly,
employing eqs(6-8), we find \begin{eqnarray} {^{(2)}R}_{0m0k} & = & -
\frac{1}{2}\partial_m\partial_k {^{(2)}g}_{00}\nonumber  \\
{^{(2)}R}_{lmnk} & = & -\frac{1}{2}(\partial_m\partial_k
{^{(2)}g}_{ln}-\partial_l\partial_k {^{(2)}g}_{mn} \nonumber \nonumber
\\ & &-
\partial_m\partial_n{^{(2)}g}_{lk}+\partial_l\partial_n{^{(2)}g}_{mk})
\nonumber \\ {^{(3)}R}_{0mnk} & =  &-\frac{1}{2}\partial_m\partial_k
{^{(3)}g}_{0n}-\frac{1}{2} \partial_n
\partial_t{^{(2)}g}_{mk}+\frac{1}{2} \partial_n\partial_m{^{(3)}g}_{k0}
\nonumber \\ & & +\frac{1}{2}\partial_k\partial_t{^2g}_{mn} \nonumber
\\ {^{(4)}R}_{0m0k} & = & -\frac{1}{2}{\partial_t}^2{^{(2)}g}_{mk}+
\frac{1}{2}\partial_k\partial_t {^{(3)}g}_{m0}+
\frac{1}{2}\partial_m\partial_t {^{(3)}g}_{k0} \nonumber \\ & &
-\frac{1}{2}\partial_m\partial_k{^{(4)}g}_{00}-
 \frac{1}{4} \partial_m{^{(2)}g}_{00}\partial_k{^{(2)}g}_{00} \nonumber
\\ & & +\frac{1}{4}\delta^{nl}\partial_n{^{(2)}g}_{00}(
\partial_m{^{(2)}g}_{lk}+ \partial_k{^{(2)}g}_{lm}-
\partial_l{^{(2)}g}_{mk}).  \end{eqnarray}

We may calculate from the above expressions the Ricci tensor
 but we will first use  the freedom to make coordinate transformations.
So  the calculation  can be done in the so-called harmonic coordinate
system which is that  one that satisfies the
 harmonic condition \[ g^{\mu\nu}\Gamma^{k}_{\mu\nu}=0 .  \] This
condition must be satisfied to all orders in the post-Newtonian
scheme.  As a result, to second and third order we get \begin{eqnarray}
\partial_n{^{(2)}g}_{00}+2\partial_m{^{(2)}g}_{mn}
-\partial_n{^{(2)}g}_{mm} & = & 0, \nonumber \\
\partial_t{^{(2)}g}_{00}+\partial_t{^{(2)}g}_{nn}-
2\partial_n{^{(3)}g}_{0n} & = &0 .  \end{eqnarray} These conditions
simplify
  considerably the expression for Ricci tensor  which can  now be
  written as \begin{eqnarray} {^{(2)}R}_{00} & = &
-\frac{1}{2}\nabla^2{^{(2)}g}_{00}, \\ {^{(2)}R}_{nm} & = &
-\frac{1}{2}\nabla^2{^{(2)}g}_{nm}, \\ {^{(3)}R}_{0n} & =  &
-\frac{1}{2}\nabla^2{^{(3)}g}_{0n}, \\ {^{(4)}R}_{00} & = &
\frac{1}{2}{\partial_t}^2{^{(2)}g}_{00}
-\frac{1}{2}\nabla^2{^{(4)}g}_{00}- \frac{1}{2}
\delta^{mk}\partial_m{^{(2)}g}_{00}\partial_k{^{(2)}g}_{00}  \\
 & & + \frac{1}{2}{^{(2)}g}_{nl}\partial_n\partial_l{^{(2)}g}_{00}.
\end{eqnarray}

We are now in a position to expand the Einstein tensor \[
G_{\mu\nu}=R_{\mu\nu}-\frac{1}{2}g_{\mu\nu}R\] as \[
G_{\mu\nu}={^{(2)}G}_{\mu\nu}+{^{(3)}G}_{\mu\nu}+... \ , \] where
$G_{00}, G_{mn}$ are of even  and $G_{0n}$ of odd order.  We may also
expand the energy-momentum tensor  in powers of $\epsilon$ as \[
T^{\mu\nu}={^{(0)}T}^{\mu\nu}+ {^{(1)}T}^{\mu\nu}+
{^{(2)}T}^{\mu\nu}+...  \] which corresponds to the $\beta=u/c$
expansion in special relativity.  As a result, the Einstein field
equations are now reduced to the system \begin{equation}
{^{(n)}G}_{\mu\nu}=\kappa {^{(n-2)}T}_{\mu\nu} \ , \end{equation}
 which can be solved step by step iteratevely. The success of this
 method is due to the fact that once the metric has been calculated  to
say $n$-th order, it can also be calculated to the next
$n\!+\!1$-order. Thus, the approximation proceeds, the complexity of
the equations increase at each step but, in principle, it can be
carried on indefinitely.

Finaly, when  we apply the post-Newtonian approximation to the case of
a  spherical symmetric object, we find that the first few terms are
actually the first terms in the expansion of the Schwarzschild metric.
However, there is no proof that the solution
 we get solving the equations in the post-Newtonian scheme converge to
the Schwarzschild one.
 The important feature of the post-Newtonian scheme is its general
applicability and its ability to give a first approximation where the
exact results are not known.

\section{The first post-Newtonian limit}

The first post-Newtonian limit of the theory is the determination of
the metric tensor to order O($\epsilon^4$) for $g_{00}$,
O($\epsilon^2$) for $g_{ij}$ and O($\epsilon^3$) for $g_{0i}$
\cite{W1,F}. Thus the metric may be written in the first post-Newtonian
limit as \[ g_{00}=-1-2U+ h_{00}+O(\epsilon^6)\ , \] \[
g_{ij}=\delta_{ij}+h_{ij}+O(\epsilon^4) \ , \] \begin{equation} g_{0i}=
h_i+O(\epsilon^5) \, \end{equation}
 where we assume that  $h_{00}, h_{ij}$ and $h_i$ are of order
 O($\epsilon^4$),  O($\epsilon^2$) and O($\epsilon^3$), respectively.
Moreover, the  Ricci tensor in eqs(11-15) may be written, to this
approximation, as \begin{eqnarray} {^2R}_{00} & = & \nabla^2
U,\nonumber \\ {^2R}_{nm} & = & -\frac{1}{2}\nabla^2 h_{nm},\nonumber
\\ {^3R}_{0n} & =  & -\frac{1}{2}\nabla^2h_n, \nonumber\\ {^4R}_{00} &
= &- {\partial_t}^2U- \frac{1}{2}\nabla^2h_{00}- 2(\nabla U)^2
\nonumber \\
 & & -h_{nl}\partial_n\partial_l U.  \end{eqnarray}

It should be noted that in general relativity, as well as in the theory
we discuss here, terms of order O($\epsilon^{2n+1}$) for $g_{00}$ and
$g_{mn}$ and O($\epsilon^{2n}$) for $g_{0n}$ are absent. This is
because these terms satisfy the  homogeneous equations \begin{equation}
\nabla^2{^{({2n+1})}g}_{00}=\nabla^2{^{({2n+1})}g}_{mn}=
\nabla^2{^{({2n})}g}_{0n}=0 \end{equation} and the only regular
solutions of the above equations are \begin{equation}
{^{({2n+1})}g}_{00}= {^{(2n+1)}g}_{00}(t) \ , {^{({2n+1})}g}_{mn}=
{^{(2n+1)}g}_{mn}(t) \ , {^{(2n)}g}_{0n}= {^{({2n})}g}_{0n}(t).
\end{equation} The fact that these solutions are unspecified functions
of time means
 that they can be eliminated by  a suitable choice of gauge.
 However, there also exists  another reason for the absence of these
terms. Under  a time-reversal transformation, the $g_{00}, g_{ij}$
components do not change sign while $g_{0i}$ do. Similarly, the
expansion parameter also changes its sign and thus from time reversal
symmetry, ignoring gravitational bremsstrahlung, the expansion in
eqs(6,17) follows immediatelly.  Furthermore, as we have seen, the
spatial derivatives are  1-order less that the time ones and this in
turn means that this approximation scheme is valid within the near zone
$r<t$.  The absence of terms of order O($\epsilon^{2n+1}$) for $g_{00}$
and $g_{mn}$ and O($\epsilon^{2n}$) for $g_{0n}$ guarantee that the
solutions of eq(16) can be matched to solutions in the far zone $r>t$
which satisfy outgoing radiation conditions at infinity \cite{Ch}.

 Let us now turn to the action in eq(1).  Denoting by (;) covariant
differentiation and ignoring gauge field contributions,  the field
equations which emerge from the
 variation of the action  are \begin{eqnarray} R_{\mu \nu}-\frac{1}{2}
g_{\mu \nu}R & = & \kappa^2 (\partial_\mu\Phi \partial_\nu
\Phi-\frac{1}{2}g_{\mu \nu}\partial_\rho \Phi \partial^\rho \Phi)
+\kappa^2 T_{\mu\nu} \nonumber \\
 &  & -\frac{\alpha'}{4} \sqrt{2} \kappa
e^{-\sqrt{2}\kappa\Phi}(R\Phi_{;\mu \nu}-R\Box\Phi g_{\mu \nu}- 2R_{\mu
\rho}{\Phi_{;\nu}}^\rho  \nonumber \\
 & & -2R_{\nu \rho}{\Phi_{;\mu}}^\rho + 2R^{\rho\sigma}\Phi_{;\rho
\sigma}g_{\mu \nu}+2R_{\mu\nu}\Box\Phi \nonumber \\ & &
-2R_{\mu\rho\nu\sigma}{\Phi_{;}}^{\rho\sigma})
+\frac{\alpha'}{2}\kappa^2e^{-\sqrt{2}\kappa\Phi} (
R\Phi_{;\mu}\Phi_{;\nu} \nonumber \\ & &
-R\Phi_{;\sigma}{\Phi_{;}}^{\sigma}g_{\mu\nu}-2R_{\mu
\rho}\Phi_{;\nu}{\Phi_;}^{\sigma}  \nonumber \\
 & & -2R_{\nu\sigma}\Phi_{;\mu}{\Phi_{;}}^{\sigma}
2R^{\rho\sigma}\Phi_{;\rho}\Phi_{;\sigma}g_{\mu\nu}+2R_{\mu\nu}\Phi_{\sigma}
{\Phi_{;}}^{\sigma} \nonumber \\ & &
-2R_{\mu\rho\nu\sigma}{\Phi_{;}}^{\rho}{\Phi_{;}}^{\sigma}) +
36\kappa^2
e^{-2\sqrt{2}\kappa\Phi}(H_{\mu\kappa\rho}{H_\nu}^{\kappa\rho}
-\frac{1}{6} g_{\mu\nu} H_{\kappa\lambda\rho} H^{\kappa\lambda\rho})
\nonumber \\ & & +12 \kappa^2(e^{-2\sqrt{2}\kappa\Phi}
 {H_\mu}^{\kappa\lambda}{R^\rho}_{\nu\kappa \lambda})_{;\rho}
 +12 \kappa^2(e^{-2\sqrt{2}\kappa\Phi}
 {H_\nu}^{\kappa\lambda}{R^\rho}_{\mu\kappa \lambda})_{;\rho} \nonumber
\\ & &  +\lambda \kappa^2
H^{\kappa\lambda\rho}J^5_\mu\varepsilon_{\kappa\lambda\rho\nu}
+\lambda\kappa^2 H^{\kappa\lambda\rho} J^5_\nu
\varepsilon_{\kappa\lambda \rho\mu}+g_{\mu\nu}\lambda\kappa^2
H_{\kappa\lambda\rho}J^5_\sigma \varepsilon^{\kappa\lambda\rho\sigma}
\nonumber \\ & & +\lambda
\kappa^2(e^{-2\sqrt{2}\kappa\Phi}g_{\lambda\mu}
\varepsilon^{\kappa\lambda\alpha\sigma}J^5_\sigma
 {R^\rho}_{\nu\alpha\kappa})_{;\rho} \nonumber \\ & & + \lambda
\kappa^2(e^{-2\sqrt{2}\kappa\Phi}g_{\lambda\nu}
\varepsilon^{\kappa\lambda\alpha\sigma}J^5_\sigma
 {R^\rho}_{\mu\alpha\kappa})_{;\rho} , \end{eqnarray} \begin{eqnarray}
\frac{1}{\sqrt{-g}}\partial_\mu(\sqrt{-g}g^{\mu\nu}\partial_\nu \Phi)
&  = & \frac{\sqrt{2}\alpha'}{16\kappa}e^{-\sqrt{2}\kappa\Phi}
R^2_{GB}-12\sqrt{2}\kappa e^{-2\sqrt{2}\kappa\Phi}H_{\mu\nu\kappa}
H^{\mu\nu\kappa} \nonumber \\ & & +\lambda \sqrt{2}\kappa
e^{-\sqrt{2}\kappa\Phi}H_{\kappa\lambda\rho} J^5_\mu
\varepsilon^{\kappa\lambda\rho\mu} , \end{eqnarray} \begin{eqnarray}
\partial_\mu \left ( \sqrt{-g} e^{-2\sqrt{2}\kappa\Phi} H^{\mu\nu\rho}
\right ) =  \frac{1}{12}\lambda \partial_\kappa \left ( \sqrt{-g}
e^{-\sqrt{2}\kappa\Phi}J^5_\mu \varepsilon^{\kappa\nu\rho\mu} \right )
.  \end{eqnarray}

The equations (21-23) above are in agreement with those, by various
field redefinitions and omissions, in refs \cite{Ax1,D,AK} For
convinience reason, let us rewrite the field equations (21) in the form
\begin{eqnarray} R_{\mu \nu} &  =  &  \kappa^2 \partial_\mu\Phi
\partial_\nu \Phi +\kappa^2 (T_{\mu\nu}-\frac{1}{2}g_{\mu\nu}T)
-\frac{\sqrt{2} \kappa \alpha'}{4}(W_{\mu\nu} -\frac{1}{2}g_{\mu\nu}W)
\nonumber \\ & & +36\kappa^2 e^{-2\sqrt{2}\kappa
\Phi}(H_{\mu\kappa\rho}{H_\nu}^{\kappa\rho} -\frac{1}{4} g_{\mu\nu}
H_{\kappa\lambda\rho}H^{\kappa\lambda\rho}) \nonumber \\ & & +12
\kappa^2(e^{-2\sqrt{2}\kappa\Phi}
 {H_\mu}^{\kappa\lambda}{R^\rho}_{\nu\kappa \lambda})_{;\rho} +12
\kappa^2(e^{-2\sqrt{2}\kappa\Phi}
 {H_\nu}^{\kappa\lambda}{R^\rho}_{\mu\kappa \lambda})_{;\rho} \nonumber
\\ & & + \lambda \kappa^2
H^{\kappa\lambda\rho}J^5_\mu\varepsilon_{\kappa\lambda\rho\nu}
 +\lambda\kappa^2 H^{\kappa\lambda\rho} J^5_\nu
 \varepsilon_{\kappa\lambda \rho\mu} \nonumber \\ & &
-2g_{\mu\nu}\lambda\kappa^2 H_{\kappa\lambda\rho}J^5_\sigma
\varepsilon^{\kappa\lambda\rho\sigma} +\lambda
\kappa^2(e^{-2\sqrt{2}\kappa\Phi}g_{\lambda\mu}
\varepsilon^{\kappa\lambda\alpha\sigma}J^5_\sigma
 {R^\rho}_{\nu\alpha\kappa})_{;\rho} \nonumber \\ & & +\lambda
\kappa^2(e^{-2\sqrt{2}\kappa\Phi}g_{\lambda\nu}
\varepsilon^{\kappa\lambda\alpha\sigma}J^5_\sigma
 {R^\rho}_{\mu\alpha\kappa})_{;\rho}, \end{eqnarray}
 where we have use the shorthand notation \begin{eqnarray} W_{\mu\nu} &
= & R(e^{-\sqrt{2}\kappa\Phi})_{;\mu \nu}-R g_{\mu\nu}\Box
e^{-\sqrt{2}\kappa\Phi}- 2R_{\mu
\rho}{(e^{-\sqrt{2}\kappa\Phi})_{;\nu}}^\rho- 2R_{\nu
\rho}{(e^{-\sqrt{2}\kappa\Phi}) _{;\mu}}^\rho + \nonumber \\
 &  & 2R^{\rho\sigma}(e^{-\sqrt{2}\kappa\Phi})_{;\rho \sigma} g_{\mu
\nu}+2R_{\mu\nu}\Box e^{-\sqrt{2}\kappa\Phi} -
2R_{\mu\rho\nu\sigma}{(e^{\sqrt{2}\kappa\Phi})_{;}}^{\rho\sigma}) .
\end{eqnarray} $W$ is the trace of the $W_{\mu\nu}$ tensor and it is
given by \begin{equation} W =
2R^{\mu\nu}(e^{f_0\kappa\Phi})_{;\mu\nu}-R\Box e^{f_0\kappa\Phi}.
\end{equation}

In order to procceed, we need also an expansion for the  antisymmetric
Kalb-Ramond field  as well as for the  dilaton $\Phi$.  The former can
be expressed as \begin{eqnarray} B_{0i} & = &
{^{(3)}B_{0i}}+{^{(5)}B_{0i}}+..., \nonumber \\ B_{ij} & = &
{^{(2)}B_{ij}}+{^{(4)}B_{ij}}+... \ .  \end{eqnarray} This particular
expansion originated from the fact that $B_{0i}$ is a vector and
$B_{ij}$ is the dual of a vector.  Thus, under parity transformations,
$B_{0i}$ changes its sing corresponding to an odd expansion in
$\epsilon$, while $B_{ij}$ do not and  corresponds to an even one.
Moreover, the field strength $H_{\mu\nu\rho}$, in this approximation,
will be given by \begin{equation}
H_{\mu\nu\rho}=\partial_{[\mu}B_{\nu\rho ]} \end{equation} since the
Cern-Simons term in eq(3) is, at least, of order $O(\epsilon^4)$.

In the same way, we may easily verify using eq(22) that the dilaton
field has also  an expansion around zero of the form \begin{equation}
\Phi= {^{(4)}\Phi}+  0(\epsilon^6) \, \end{equation} where the first
term ${^{(2)}\Phi}$ is absent in view of the lack of matter sources for
the dilaton.

Let us now turn to the tensor $W_{\mu\nu}$ defined above.
 The expansion of this tensor will look like \[
W_{\mu\nu}={^{(6)}W}_{\mu\nu}+{^{(7)}W}_{\mu\nu}+ ...  \]as follows
from eq(9,17,18,29) and the definition (25).  Explicitly, we find that
its first non-vanishing component  is \begin{eqnarray}
{^{(6)}W}_{00}&=&-\sqrt{2}\kappa( ({^{(2)}R}_{00}+{^{(2)}R}^n_n)
\nabla^2{^{(4)}\Phi} \nonumber \\
 & & -2{^{(2)}R}^{mn}\partial_m\partial_n{^{(4)}\Phi}-2{^{(2)}R}^{0m0n}
\partial_m\partial_n{^{(4)}\Phi} ).  \end{eqnarray} Thus, although
$W_{\mu\nu}$ will contribute to the gravitational field, this
contribution will start operating only at the post post-Newtonian limit
since the leading term in the $\epsilon$-expansion of $W_{\mu\nu}$ is
of order $\epsilon^6$.  Similarly, the Gauss-Bonnet term (2) is given
to this approximation by \begin{eqnarray} R^2_{GB} & = & 8
(\partial_m\partial_kU)(\partial^m\partial^kU) -8(\nabla^2U)^2
\nonumber \\
 & = & 8 \partial_m ( \partial_kU \partial^m\partial^kU- \partial^mU
\nabla^2U).  \end{eqnarray} Using eqs(17,18,27-31) and units
$\kappa^2=8\pi$, the field equations may now be written as
\begin{eqnarray} \nabla^2 U & = & 4\pi  {^{(0)}T}^{00},  \\ \nabla^2
h_n & = & 16\pi {{^{(1)}T}}^{n0},  \\ \nabla^2 h_{mn} & = & -8\pi
\delta_{mn} {^{(0)}T}^{00},  \\ \nabla^2 h_{00} & = & -2\partial_t^2
U-2\nabla^2 U^2-8\pi G
 ({^{(2)}T}^{00}+{^{(2)}T}^n_n),   \\ \nabla^2{^{(4)}\Phi} & =&
\frac{\sqrt{2}\alpha'}{16\kappa}
(8(\partial_m\partial_kU)(\partial^m\partial^kU)
-8(\nabla^2U)^2)-12\sqrt{2}\kappa H_{\mu\nu\lambda}H^{\mu\nu\lambda},
\\ \partial_nH^{n0m} & = & \frac{1}{12} \partial_n J^5_l
\varepsilon^{n0ml} \\ \partial_n H^{nml} & =& 0.  \end{eqnarray}

We observe that the first three equations  above  which specifies the
metric are  identical  to the post-Newtonian evaluation of the field
equations in  general  relativity.  Thus, there is no way to
distinguish, to this approximation,
 Einstein and superstring gravity.

The  system of equations (32-38) form a consistent system and if the
energy-momentum tensor is given, we can specify the graviton, the
dilaton and the axion field to the first post-Newtonian limit.  We will
illustrate this below in two specific examples, namely, we will try to
find the fields of a spinning spherical symmetric mass as well as of an
infinite long  cosmic string.

To begin with the former case, let us note that the energy-momentum
tensor for the spinning point mass is given by \begin{eqnarray}
{^{(0)}T}^{00} & = & m\delta^{(3)}({\bf r}), \\ {^{(1)}T}^{0n} & = &
-\frac{1}{2}\varepsilon^{nmk} s_m \partial_k\delta^{(3)}({\bf r})
\end{eqnarray} where \begin{eqnarray} m & = & \int d^3x {^{(0)}T}^{00},
\\ s_n & = & \varepsilon_{nmk} \int d^3x x^m {^{(1)}T}^{0k}
\end{eqnarray} are the mass and the spin ${\bf s}=(s_n)$  of the point
particle, respectively. Moreover the spin density is given by \[ S_n=
s_n \delta^{(3)}({\bf r}).  \]
 The energy-momentum tensor (39,40) specifies the  metric to be
\begin{eqnarray} U({\bf x},t) & = & - \int \frac{{^{(0)}T}^{00} ({\bf
x'},t) d^3x}{|{\bf x}-{\bf x'}|}=-\frac{m}{r} \ , \\ h_n({\bf x},t) & =
& -4 \int
 \frac{{^{(1)}T}^{0n}({\bf x'},t) d^3x}{|{\bf x}-{\bf x'}|}=
-2\varepsilon_{nmk}\frac{s^mx^k }{r^3} \ ,  \\ h_{00}(x,t)&  =  &
-\frac{2m^2}{r^2} +O(r^{-4}) \ .  \end{eqnarray}

Thus, the metric for a spinning point mass in spherical coordinates
with the direction ${\bf S}$ as the symmetry axis may be written as
\begin{eqnarray} ds^2 & = & -\left( 1-\frac{2m}{r}+\frac{2m^2}{r^2}
  \right)dt^2 -4\varepsilon_{ijk}\frac{S^j x^k}{r^3}dt dx^i \nonumber
\\ & & +\left(1+\frac{2m}{r}\right) \left(dr^2+r^2d\theta^2
+r^2sin^2\theta d\phi^2\right) \nonumber.  \end{eqnarray} which is, as
expected,  the first post-Newtonian limit of the Kerr metric.

 Let us now turn to the antisymmetric two-form field.  In view of the
invariance of $H_{\mu\nu\rho}$ in eq(28) under the gauge transformation
\begin{equation} B_{\mu\nu} \rightarrow B_{\mu\nu}+\partial_\mu
\Lambda_\nu-\partial_\nu \Lambda_\mu, \end{equation} where
$\Lambda_\mu$ are the components of an  one-form, we may fix the gauge
freedom by imposing the  Coulomb condition \begin{equation}
\partial_nB^{0n}={\bf \nabla B}=0.  \end{equation} In this particular
gauge,  eqs(37,38) reduce to \begin{eqnarray} \nabla^2 {\bf B} &
=&\frac{1}{2}\nabla \times {\bf S} \\ H_{nml}&=&0 \end{eqnarray} where
we have used the notation \[ {\bf B}=({^{(3)}B}_{0n}) \] and the fact
that the axial current is  given by \[ J^5_\mu=(0,{\bf S}) \] with
${\bf S}$ the  spin density.  One may now  easily  verify that the
solution to eq(48) may be written as \begin{eqnarray} {\bf B} & =&
\frac{\lambda}{8\pi} \frac{{\bf s \times x}}{r^3} \end{eqnarray}
Moreover, as follows from eq(49), the space-space components of the
antisymmetric two-form field  are constants and thus \begin{equation}
B_{mn}={^{(2)}B}_{mn}=0 \end{equation} by a gauge transformation.
Finally, one has to verify  for consistency reasons that the solutions
(43-45,50) satisfy the gauge conditions (10,47) as well. It is not
difficult to prove that these conditions are indeed satisfied, (47)
automatically and (10) by virtue of \[ \nabla \cdot {\bf h}=0 \] where
${\bf h}$ is the vector with components $(h_1,h_2,h_3)$.

 The post-Newtonian scheme can  applied  to infinitely extended objects
as well. For example, let us examine the
 case of an infinitely long in the z-direction cosmic string which is
characterized by the equation of state \begin{equation}
T^{00}+T^{zz}=0.  \end{equation} Moreover, we will assume that it is
rotating so that its energy momentum tensor may be written as :
\begin{eqnarray} {^{(0)}T}^{00} & = & \mu \delta^{(2)}({\bf
r}),\nonumber \\ {^{(0)}T}^{zz} & = & -\mu \delta^{(2)}({\bf r}),
\nonumber \\ {^{(0)}T}^{ij} & = & 0, \nonumber \\ {^{(1)}T}^{0i} &  =
& -\frac{1}{2}\varepsilon^{ij}\sigma \partial_j \delta^{(2)}({\bf r})
\end{eqnarray} where $i,j=1,2$ and $ \mu$ and $\sigma$ are the mass and
the spin density, respectively.

 The Einstein equations (16) are written to lowest order as :
\begin{eqnarray} {^{(2)}R}_{00} & = & \kappa
 \left( {^{(0)}T}_{00}+\frac{1}{2}(-{^{(0)}T}_{00}
+{^{(0)}T}_{zz})\right), \nonumber \\ {^{(2)}R}_{ij} & = &
-\frac{\kappa}{2}\delta_{ij}(-{^{(0)}T}_{00} +{^{(0)}T}_{zz}),
\nonumber \\ {^{(2)}R}_{zz} & = & \kappa
 \left(
 {^{(0)}T}_{zz}-\frac{1}{2}(-{^{(0)}T}_{00}+{^{(zz)}T}_{zz})\right),
 \nonumber \\ {^{(3)}R}_{0i} & = &  \kappa {^{(1)}T}_{0i}.
\end{eqnarray} Employing eqs(11--13,17,53) and $\kappa^2=8\pi$ we get
\begin{eqnarray} \nabla^2 U & = & 0,   \nonumber \\ \nabla^2 h_{ij} &
=  & -16\pi \mu \delta^{(2)}({\bf r}) \delta_{ij},
 \nonumber \\ \nabla^2 h_{zz} & = &  0, \nonumber \\ \nabla^2h_i & = &
-8\pi \epsilon_{ij} \sigma \partial_j\delta^{(2)}({\bf r }) .
\end{eqnarray}

The solution of this system is \begin{eqnarray} U(\rho) & = & const.,
\nonumber \\ h_{ij} & = & -8\mu \delta_{ij}
\ln\left(\frac{\rho}{\rho_0}\right), \nonumber \\ h_{zz} & = & const.,
\nonumber\\ h_i({\bf r}) & = & -4\sigma \frac{\varepsilon^{ij}
x_j}{\rho^2}.  \end{eqnarray} For the calculation of $h_i({\bf r})$
the ansatz \[ h_i({\bf r})=\varepsilon^{ij} \partial_j \Lambda ({\bf
r}) + \partial_i K({\bf r}).  \] have been employed. As a result, the
metric in this approximation can be written as \[ ds^2 =- dt^2 + dz^2
+\left(1-4\pi \ln(\rho/\rho_0)\right)(d\rho^2+\rho^2d\phi^2) +
\frac{8\sigma}{\rho}dt d\phi.  \] This is the exact metric \cite{DJ},
in the $\sigma=0$ limit, of a cosmic string with deficit angle \[
\Delta\phi =\pi(1+4\mu).  \]

The eq(48) now for the antisymmetric field can be written as \[
\nabla^2 B_i({\bf r})=\frac{1}{2}\sigma \varepsilon^{ij}\partial_j
\delta^{(2)}({\bf r}) \] and employing a similar to (57) ansatz, we
find that \begin{equation} B_i({\bf r})= \frac{\sigma}{4\pi}
\varepsilon^{ij} \frac{x_j}{\rho^2}.  \end{equation}

The corresponding $H_{\mu\nu\rho}$ field is zero and the solution (58)
is the counterpart of the solution in \cite{Ax2} in the cylidrical
symmetric case.

\section{Radiation}

Another approximation method in general relativity is the weak-field
approximation where small perturbations (graviton modes) of the
Minkowski metric \cite{W1,F} are considered. The field equations are
reduced then to the wave equation with source the energy-momentum
tensor and thus, in this case,  propagating gravitational modes
emerge.  Today, the  analysis of almost 20 years observational data
from the binary pulsar PRS 1913+16 \cite{BP} seems to prove the
existence of gravitational radiation \cite{T} and it remains a direct
detection of gravitational waves (by means of a gravitational antenna)
\cite{Sz}.
 We will examine here wave and radiation aspects of the superstring
gravity as follows from eqs(21-23).

To begin with, let us recall that in the wave-analysis it is convinient
to introduce the quasi-orthonormal
 null-tetrad base\cite{NP} \begin{eqnarray} k& =&
\frac{1}{\sqrt{2}}\left( \hat e_t+\hat e_z \right),\nonumber \\ \ell &
=& \frac{1}{\sqrt{2}}\left( \hat e_t-\hat e_z \right), \nonumber \\ m&
=& \frac{1}{\sqrt{2}}\left( \hat e_x+i\hat e_y \right), \nonumber \\
\overline m& =& \frac{1}{\sqrt{2}}\left( \hat e_x-i\hat e_y \right),
\end{eqnarray} where the tetrad vectors satisfy the relation \[ m \cdot
\overline m = -k\cdot \ell =1.  \] Let us consider for example  the
electromagnetic field.
 The measurable quantity is the field strengh $F_{\mu\nu}$. For plane
waves travelling in the +z direction, $F_{\mu\nu}$ is a function of
the``retarded" time u=t-z.  The components of the field strengh in the
null-tetrad base are $F_{\ell p}, F_{pq}$, where $p,q=k,m,\overline m$.
The Bianchi identity for $F_{\mu\nu}$ is written as \[ \partial_q
F_{\ell p}+\partial_p F_{q\ell}+\partial_{\ell} F_{pq}=0, \] or \[
\partial_{\ell}F_{pq}=0 \]
 where $\partial_p, \partial_\ell$  denote derivatives to the $p,\ell$
 direction.  Thus, the components $F_{pq}$ are constants and there
exist three independent components $F_{\ell p}$ which correspond to the
three possible  polarizations of the photon.

If moreover $F_{\mu\nu}$ satisfies Maxwell equations \[ \partial_\mu
F^{\mu \nu}=0, \] then in the basis (59) these equations are written as
\[ \partial_{\ell}F^{\ell k}=0 \] and thus $F_{\ell k}$, like
$F_{pq}$,  is  constant (non propagating).
 As a result, Maxwell equations specify only the $F_{\ell k}$ component
and there are finally two independent modes
 $F_{\ell m}, F_{\ell\overline m}$ to carry  the electromagnetic
 energy.  These components are the two polarizations modes of the
photon.

One may  now proceed in the same lines with the gravitational field.
The measurable quantity here is the Riemann tensor.
 There is an elegant way to analyze the polarization of gravitational
waves
 and it is provided by the Newman-Pernose (NP) formalism \cite{NP}.
For the Riemann tensor, there are six algebraically  independent
components which correspond to the six possible polarizations  of the
gravitational waves. These are, in the NP formalism, the two real
functions $\Psi_2(u)$ and $\Phi_{22}(u)$ and the two complex functions
$\Psi_3(u)$, $\Psi_4(u)$.  These functions are actually the components
\begin{eqnarray} \Psi_2 & = & -\frac{1}{6}R_{\ell k\ell k}, \nonumber
\\ \Psi_3 & = & -\frac{1}{2}R_{\ell k \ell \overline m}, \nonumber \\
\Psi_4 & = & -R_{\ell \overline m\ell \overline m}, \nonumber \\
\Phi_{22} & = & -R_{\ell m\ell \overline m}, \end{eqnarray} of the
Riemann tensor in the null-tertad base (59).  To find the independent
polarization modes for the graviton, one considers the vacuum field
equations (no matter sources) and keeps only linear terms in the
fields. One may then verify that the equations for the graviton is
exactly the same as in general relativity and lead to \[
\Psi_2=\Psi_3=\Phi_{22}=0 \ , \   \Psi_4 \neq 0 \] Thus,  there exist
two independent modes corresponding to the two possible polarizations
of the graviton. This is not the case in alternative  theories  for
gravity. In Brans-Dicke theory for example, in addition to $\Psi_4 \neq
0$, we also have $\Phi_{22} \neq 0$ \cite{ELL}.

Let us now proceed with the gravitational radiation of the superstring
effective action in the dominant $O(\alpha'^0)$ order.   Here the
theory contains the graviton, the dilaton and the antisymmetric
two-form field, with $H_{\mu\nu\rho}$ given in eq(28), and the field
equations may be written as \begin{eqnarray}
R_{\mu\nu}-\frac{1}{2}g_{\mu\nu}R & = & \kappa^2 T_{\mu\nu} +
36\kappa^2 e^{-2\sqrt{2}\kappa \Phi}
(H_{\mu\kappa\lambda}{H_\nu}^{\kappa\lambda}
-\frac{1}{6}g_{\mu\nu}H_{\kappa\lambda\rho}H^{\kappa\lambda\rho})
\nonumber \\ & & +\kappa^2(\partial_\mu \Phi \partial_\nu
\Phi-\frac{1}{2} g_{\mu\nu}\partial_\rho \Phi \partial^\rho \Phi)+
\lambda \kappa^2
H^{\kappa\lambda\rho}J^5_\mu\varepsilon_{\kappa\lambda\rho\nu}
\nonumber \\ & & +\lambda\kappa^2 H^{\kappa\lambda\rho} J^5_\nu
\varepsilon_{\kappa\lambda \rho\mu}+g_{\mu\nu}\lambda\kappa^2
H_{\kappa\lambda\rho}J^5_\sigma \varepsilon^{\kappa\lambda\rho\sigma},
\end{eqnarray} \begin{eqnarray} \partial_\mu (\sqrt{-g}
e^{-2\sqrt{2}\kappa\Phi} H^{\mu\nu\rho} ) =  \frac{1}{12}\lambda
\partial_\kappa (\sqrt{-g} e^{-\sqrt{2}\kappa\Phi}J^5_\mu
\varepsilon^{\kappa\nu\rho\mu}) \ , \end{eqnarray} \begin{eqnarray}
\frac{1}{\sqrt{-g}}\partial_\mu (\sqrt{-g} g^{\mu\nu} \partial_\nu
\Phi) &= & -12\sqrt{2}\kappa
e^{-2\sqrt{2}\kappa\Phi}H_{\mu\nu\rho}H^{\mu\nu\rho} +\lambda \sqrt{2}
\kappa H_{\kappa\lambda\rho}
 J^5_\mu \varepsilon^{\kappa\lambda\rho\mu}.  \end{eqnarray} We will
evaluate the above field equations in the  weak-field approximation and
we will neglect the dilaton field in view of its $\kappa^2$ coupling to
its sources in eq(63). In particular, we will consider small
perturbations $h_{\mu\nu}$ around the vacuum Minkowski metric
$\eta_{\mu\nu}$.  We may define
 a tensor $\theta_{\mu\nu}$ by \[
\theta_{\mu\nu}=h_{\mu\nu}-\frac{1}{2}\eta_{\mu\nu}h^\lambda_\lambda,
\] so that the field equations in the gauge \begin{equation}
\partial^\mu \theta_{\mu\nu}=0 \end{equation} and in the linearized
limit can be written as \begin{equation} \Box
\theta_{\mu\nu}=-2\kappa^2(T_{\mu\nu}+t_{\mu\nu}).  \end{equation}
 $t_{\mu\nu}$ is quadratic in the fields and it is given by
\begin{eqnarray} t_{\mu\nu} & =&
-\frac{1}{\kappa^2}({R^{(2)}}_{\mu\nu}-\frac{1}{2}
\eta_{\mu\nu}R^{(2)}) \nonumber \\ & &
+36(H_{\mu\kappa\lambda}{H_\nu}^{\kappa\lambda}
-\frac{1}{6}\eta_{\mu\nu}H_{\kappa\lambda\rho}H^{\kappa\lambda\rho})
\end{eqnarray} with ${R^{(2)}}_{\mu\nu}$ the quadratic  in $h_{\mu\nu}$
part of the Ricci tensor. Similarly, the equations for the
antisymmetric two-form field may be written as \[ \Box
B_{\nu\rho}+\partial^\mu \partial_\rho B_{\mu\nu}+\partial^\mu
\partial_ \nu B_{\rho\mu}= \frac{1}{2}\lambda \partial^\kappa J^{5\mu}
\varepsilon_{\kappa \nu\rho\mu}.  \]
 The gauge freedom in the definition of $H_{\mu\nu\rho}$ in eq(28),
 allows one to consider a   particular  gauge system. Thus, we may
 choose the Lorentz gauge \begin{equation} \partial^\mu B_{\mu\nu}=0,
\end{equation} in which the B-field satisfies the wave equation
\begin{equation} \Box B_{\nu\rho} = \frac{1}{2}\lambda \partial^\kappa
J^{5\mu} \varepsilon_{\kappa \nu\rho\mu}.  \end{equation} We are
interested now in plane wave solutions in the  radiation zone of the
form \begin{eqnarray} h_{\mu\nu}& = & e_{\mu\nu}e^{i({\bf k \cdot
x}-\omega t)}+ e^*_{\mu\nu}e^{-i({\bf k \cdot x}-\omega t)} , \\
B_{\mu\nu}& = & \epsilon_{\mu\nu}e^{i({\bf q \cdot x}-\omega_q t)}+
\epsilon^*_{\mu\nu}e^{-i({\bf q \cdot x}-\omega_q t)} , \end{eqnarray}
where $e_{\mu\nu}$ is symmetric and $\epsilon_{\mu\nu}$ is
antisymmetric. Moreover, $k_\mu=(\omega,{\bf k})$ and
$q_\mu=(\omega_q,{\bf q})$ are  the null wave 4-vectors for the
graviton and the antisymmetric field, respectively, so that
\begin{equation} k^\mu k_\mu = q^\mu q_\mu =0.  \end{equation} The
gauge conditions (64,67) are written for the solutions (69,70) as
\begin{eqnarray} k^\mu e_{\mu\nu}& =& 0, \\ q^\mu \epsilon_{\mu\nu}&= &
0.  \end{eqnarray} Employing the expressions (69-73)  in (66) and
integrating
 over spacetime, we may write \begin{equation} <t_{\mu\nu}>=
\frac{k_\mu k_\nu}{\kappa^2} ({e^*}_{\kappa\lambda}e^{\kappa
\lambda}-\frac{1}{2}|e^\lambda_\lambda|^2)+ 72q_\mu q_\nu
{\epsilon^*}_{\kappa \lambda} \epsilon^{\kappa\lambda}.  \end{equation}
The brackets denote average over several wavelengths or over spacetime
regions large compared to the characteristic wavelength.
 For the graviton, as we have seen,  there are two independent modes
which are  actually
 the transverse-traceless components $\theta^{ij}_{TT}=h^{ij}_{TT}$  of
the metric,
 defined in terms of the projection operator \[ P^i_j=\delta^i_j-n^i
n_j, \] with $n^i=\frac{{\bf x}^i}{r}$, as \begin{equation}
\theta^{ij}_{TT}=P^i_k \theta^{kl} P^j_l
 -\frac{1}{2} P_{kl} \theta^{kl} P^{ij}